\documentclass[twocolumn,aps,pra,preprintnumbers,bibnotes10pt,superscriptaddress]{revtex4}

\usepackage{amsmath}
\usepackage{empheq}
\usepackage{graphicx}
\usepackage{ulem}
\usepackage{dcolumn}
\usepackage{epsfig}
\usepackage{bm}
\usepackage{array}
\usepackage[english]{babel}
\usepackage{hyperref}
\hypersetup{
colorlinks=true,
citecolor=black,
linkcolor=black,
urlcolor=black
 , bookmarks=true,
 pdfmenubar=true
}

\usepackage{graphicx}
\usepackage{amsmath}
\usepackage{bm}           
\usepackage{bbm}
\usepackage{ulem}
\usepackage{appendix}

\newcommand*\widefbox[1]{\fbox{\hspace{2em}#1\hspace{2em}}}

\newcommand{\Eins}{\ensuremath{\mathbbm 1}}

\newcommand{\vect}[1]{\bm{#1}}

\newcommand{\be}{\begin{equation}}
\newcommand{\ee}{\end{equation}}

\newcommand{\beq}{\begin{eqnarray}}
\newcommand{\eeq}{\end{eqnarray}}

\newcommand{\psis}{\psi_s}
\newcommand{\vtp}{\vect{\varphi}}

\makeatletter

\newcommand{\vts}{\vect{\theta}}

\begin{document}

\title{
Optimal measurements for simultaneous quantum estimation of multiple phases
}

\author{Luca Pezz\`e}
\affiliation{QSTAR, INO-CNR and LENS, Largo Enrico Fermi 2, I-50125 Firenze, Italy}

\author{Mario A. Ciampini}
\affiliation{Dipartimento di Fisica, Sapienza Universit\`{a} di Roma,
Piazzale Aldo Moro 5, I-00185 Roma, Italy}

\author{Nicol\`o Spagnolo}
\affiliation{Dipartimento di Fisica, Sapienza Universit\`{a} di Roma,
Piazzale Aldo Moro 5, I-00185 Roma, Italy}

\author{Peter C. Humphreys}
\affiliation{Department of Physics, Clarendon Laboratory, University of Oxford, Oxford  OX1 3PU, United Kingdom }

\author{Animesh Datta}
\affiliation{Department of Physics,  University of Warwick, Coventry CV4 7AL, United Kingdom }

\author{Ian A. Walmsley}
\affiliation{Department of Physics, Clarendon Laboratory, University of Oxford, Oxford  OX1 3PU, United Kingdom }

\author{Marco Barbieri}
\affiliation{Dipartimento di Scienze, Universit\`a degli Studi Roma Tre, Via della Vasca Navale 84, 00146, Rome, Italy }

\author{Fabio Sciarrino}
\affiliation{Dipartimento di Fisica, Sapienza Universit\`{a} di Roma,
Piazzale Aldo Moro 5, I-00185 Roma, Italy}

\author{Augusto Smerzi}
\affiliation{QSTAR, INO-CNR and LENS, Largo Enrico Fermi 2, I-50125 Firenze, Italy}

\begin{abstract}
A quantum theory of multiphase estimation is crucial for quantum-enhanced sensing and imaging and
may link quantum metrology to more complex quantum computation and communication protocols.
In this letter we tackle one of the key difficulties of multiphase estimation: obtaining a measurement which saturates the
fundamental sensitivity bounds. We derive necessary and sufficient conditions for projective measurements acting on pure states to saturate the maximal theoretical bound on precision given by the quantum Fisher information matrix. We apply our theory to the specific example of interferometric phase estimation using photon number measurements,
a convenient choice in the laboratory. Our results thus introduce concepts and methods relevant to the future theoretical and experimental development of multiparameter estimation.
\end{abstract}

\date{\today}

\maketitle

{\it Introduction.}
Quantum metrology is currently attracting considerable interest in the light of its technological applications.
Theoretical developments and experimental investigations have, so far, mostly focussed on the estimation 
of single phase \cite{GiovannettiNATPHOT2011, PezzeRMP2017,Paris2009}, for which the
ultimate sensitivity bounds and explicit conditions for their saturation are well known \cite{BraunsteinPRL1994, BraunsteinANNPHYS1996}.
These studies have been further extended in order to understand the connection between enhancement in 
phase estimation and particle entanglement~\cite{GiovannettiPRL2006, PezzePRL2009, HyllusPRA2012, PezzePNAS2016}, 
as well as the impact of noise and dissipation on the fundamental bounds~\cite{DobrzanskiNATCOMM2012, EscherNATPHYS2012}. 
Several proof-of-principle experiments have demonstrated phase estimation below the classical (shot-noise) limit \cite{PezzeRMP2017}, 
including applications in fields as diverse as magnetometry \cite{OckeloenPRL2013}, atomic clocks \cite{Louchet-ChauvetNJP2010} and 
optical detection of gravitational waves \cite{Ligo2011}.

Yet, a significant class of problems can not be efficiently cast as the estimation of a single parameter, as is the case for quantum sensing and imaging~\cite{Preza99}, and for quantum communication and computation protocols \cite{Granade2012,Huszar2012}. Such multiparameter cases have been the subject of recent efforts, investigating the role of entanglement \cite{Ciampini2015}, and 
the impact of noise and decoherence \cite{YueSR2014,Knott2016}. Explicit examples have been considered, including measurement strategies for state estimation \cite{Gill2000,Li2016}, the joint estimation of phase and loss rate~\cite{Piniel,CrowleyPRA2014}, phase and phase diffusion~\cite{VidrighinNATCOMM2014,Knysh2013,Altorio2015,Szcz2017}, components of a displacement in phase space \cite{Genoni,Schnabel}, multiple phases~\cite{HumphreyPRL2013,Ciampini2015}, parameters belonging to multidimensional fields \cite{Baumgratz2016}, and estimation tasks with partial knowledge on the actual measurement device \cite{Altorio2016}. 

However, there are still several open questions in multiparameter estimation, one of the most urgent of which is to find saturable lower bounds of phase sensitivity. 
There exists a fundamental bound -- the quantum Cramer-Rao bound -- which has been formulated in~\cite{Helstrom} for the multiparameter case. 
However, unlike in the single-parameter case, the quantum Cram\'er-Rao bound is not always saturable, with a necessary condition provided 
by the commutativity of the symmetric logarithmic derivatives. Furthermore, it does not provide a recipe for constructing optimal 
measurements \cite{Paris2009,Helstrom, MatsumotoJPA2002, HumphreyPRL2013, RagyPRA2016}. 

In this manuscript, we discuss the properties that a projective measurement must have in order to saturate the quantum Cramer-Rao bound 
for multiphase estimation with probe pure states. 
Our results extend and complement previous works by Helstrom~\cite{Helstrom}, Matsumoto \cite{MatsumotoJPA2002}, and Humphreys et al.~\cite{HumphreyPRL2013}, 
by identifying the necessary and sufficient conditions on the projectors. 
Differently from earlier investigations, we do not focus on specific instances, but provide general constructive conditions for obtaining an optimal 
measurement not limited by non-commutativity. This has implications for the future experimental and theoretical development of multiphase estimation 
with quantum technologies, including photons, atoms and trapped ions.


%
\begin{figure}[b]
\centering
\includegraphics[width=0.5\textwidth]{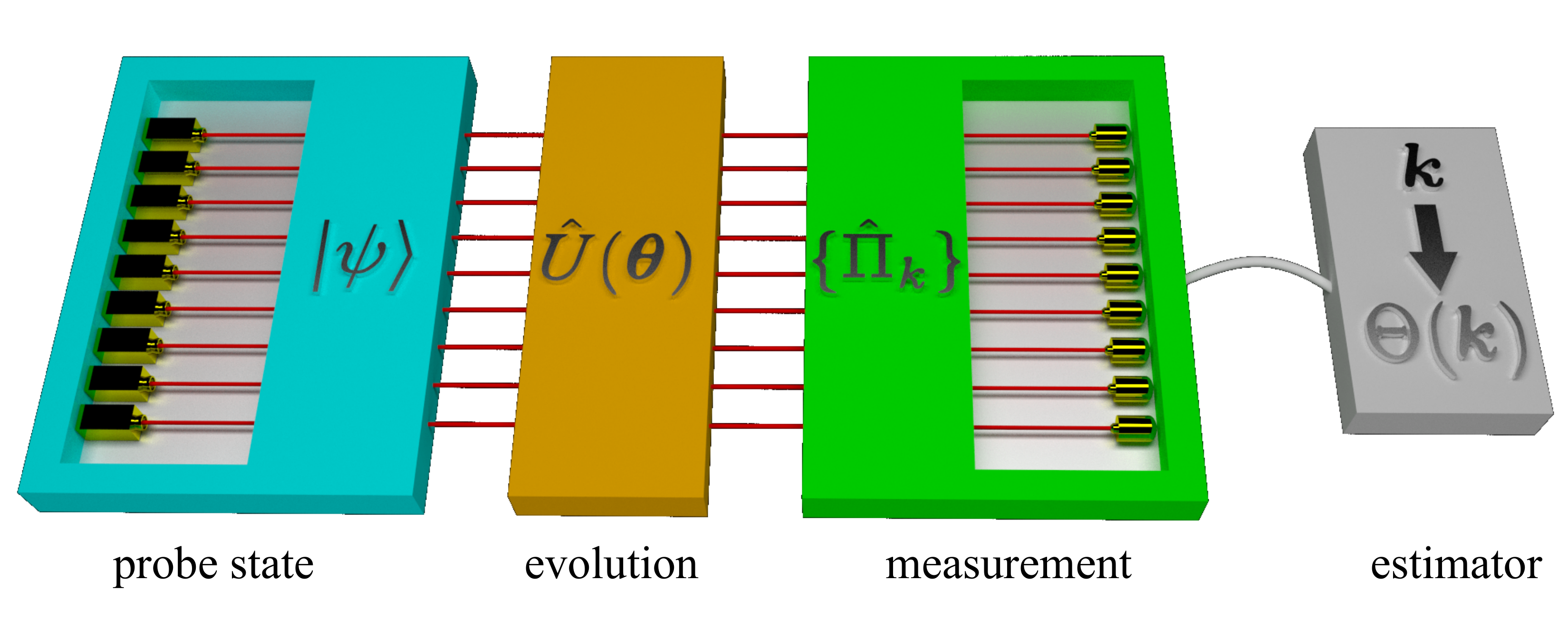}
\caption{General framework of multiparameter estimation considered in this manuscript: 
a probe state $\vert \psi \rangle$ is prepared, transformed according to the unitary
parameter-dependent transformation $\hat{U}(\vts)$ and measured by a set of projectors $\{ \hat{\Pi}_k\}$. 
The value of the vector parameter is retrieved via an estimator $\Theta(\vect{k})$, function of the measurement outcomes.}
\label{fig:multipar}
\end{figure}
%


{\it Multiphase estimation.}
The simultaneous estimation of a $d$-dimensional vector parameter $\vts=\{\theta_1, \theta_2, ..., \theta_d \}$
follows the standard steps of quantum metrology (see Fig. \ref{fig:multipar}):
{\it (i)} A probe state is prepared (here we will consider a pure probe state $\vert \psi \rangle$).
{\it (ii)} It is shifted in phase by applying a phase-encoding unitary transformation $\hat{U}_{\vts}$,
the output state being $\vert \psis \rangle \equiv \hat{U}_{\vts} \vert \psi \rangle$.
{\it (iii)} The output state is detected.
In the following we will consider a set of projective measurements $\{ \hat \Pi_k \}_k$,
labelled by the index $k$ representing the possible result.
Eventually, the protocol is repeated $\nu$ times using independent and identical copies of the output state
(with the same transformation and output measurement). 
{\it (iv)} Finally, from the output results $\vect{k} \equiv \{ k_1, ..., k_\nu \}$ one infers the vector parameter
via a suitably chosen function $\Theta(\vect{k})$, called the estimator.
The probability of observing the sequence $\vect{k}$, conditioned to the vector parameter $\vts$, 
is given by $P(\vect{k} \vert \vts) = \prod_{i=1}^\nu P(k_i \vert \vts)$
and $P(k \vert \vts) = \langle \psis  \vert \hat{\Pi}_k \vert \psis \rangle$.
In the following we will consider locally unbiased estimators, for which the average value of the
estimator equals the true value of the parameter:
$\bar{\Theta}(\vect{k}) = \sum_{\vect{k}}  P(\vect{k} \vert \vts) \Theta(\vect{k}) =  \vts$
and $d\bar{\Theta}(\vect{k})/d\theta_l=1$ ($l=1,...,d$).
A figure of merit of phase sensitivity is the covariance matrix,
\be 
[\vect{B}(\vect{\theta})]_{l,m} = \sum_{\vect{k}} P(\vect{k} \vert \vts) \big(\bar{\vts} - \vts\big)_l \big(\bar{\vts} - \vts\big)_m.
\ee
The diagonal elements $\vect{B}_{l,l}$ equal the variance $(\Delta \theta_l)^2$.
For any unbiased estimators, and independent measurements, the chain of inequalities
\be
\label{bound}
\vect{B}(\vect{\theta}) \geq \frac{[\vect{F}(\vect{\theta})]^{-1} }{\nu} \geq \frac{[\vect{F}_Q(\vect{\theta})]^{-1} }{\nu}
\ee
holds (in the matrix sense). The first inequality is the Cramer-Rao bound, where
\be
[\vect{F}(\vect{\theta})]_{l,m}
= \sum_k \frac{\partial_l P(k \vert \vts) \, \partial_m P(k \vert \vts)}{P(k \vert \vts)}
\ee
is the $d \times d$ symmetric Fisher information matrix (FIM), and $\partial_{l} = \partial/\partial \theta_{l}$. 
Since $\partial_l P(k \vert \vts) = 2 {\rm Re}[\langle \partial_l \psis  \vert \hat{\Pi}_k \vert \psis \rangle]$,
the FIM depends on how the measurement set acts on the Hilbert subspace $\mathcal{H}$ spanned by
the probe state $\vert \psis \rangle$ and 
 $\{ \vert \partial_l \psis \rangle \}_{l=1,...,d}$.
Here ${\rm Re}[x]$ and ${\rm Im}[x]$ indicate the real and imaginary part of $x$, respectively,
and $\vert \partial_l \psis \rangle = -i \hat{H}_l \vert \psis \rangle$, where
$\hat{H}_l \equiv i [\partial_l \hat{U}(\vts)] \hat{U}^{\dag}(\vts)$ is a Hermitian operator.
The Cramer-Rao bound can be established only if the FIM is strictly positive (and thus invertible).
In this case, it can always be saturated, asymptotically in $\nu$, by the maximum likelihood estimator \cite{Helstrom}.
The second bound in Eq.~(\ref{bound}) is due to $\vect{F}(\vect{\theta}) \leq \vect{F}_Q(\vect{\theta})$ \cite{Helstrom},
where $\vect{F}_Q(\vect{\theta})$ is the quantum Fisher information matrix (QFIM),
\be \label{QuantumFisher}
[ \vect{F_Q}(\vect{\theta}) ]_{l,m}  = 4 {\rm Re}[
\langle \partial_l \psis \vert \partial_m \psis \rangle ]
+ 4
\langle \partial_l \psis \vert \psis \rangle
\langle \partial_m \psis \vert \psis \rangle.
\ee
We recall that the inequality $\vect{F} \leq \vect{F_Q}$ is understood in the matrix sense, {\it i.e.} 
$\vect{u}^\top \vect{F} \vect{u} \leq \vect{u}^\top \vect{F_Q} \vect{u}$ holds for 
arbitrary $d$-dimensional real vectors $\vect{u}$. 
The inverse of the QFIM -- when it exists -- sets a lower bound (\ref{bound}) for the simultaneous estimation of multiple parameters,
called the quantum Cramer-Rao bound, which only depends on the probe state and
phase encoding transformation.
The QFIM is a particularly appealing quantity:
it is proportional to the second-order Taylor expansion of the multidimensional Bures distance
and, equivalently, the square root of the fidelity between quantum states~\cite{BraunsteinPRL1994}.
It is a quantum statistical speed quantifying how much the output state differs from the input one when applying small phase shifts~\cite{PezzePNAS2016, StrobelSCIENCE2014}.
There is a major problem though: while in the single parameter case it is always possible to choose an optimal measurement
for which the equality $\vect{F}{=}\vect{F_Q}$ holds~\cite{BraunsteinPRL1994, BraunsteinANNPHYS1996}, in the multiparameter case ($d>1$)
such an optimal measurement does not exist in general.

The search for conditions under which the FIM saturates the QFIM has long engaged the field of quantum metrology~\cite{Helstrom, MatsumotoJPA2002}.
For non-commuting operators the main result available in the literature is due to Matsumoto~\cite{MatsumotoJPA2002}: 

{\it Weak commutativity theorem (Matsumoto).} 
Given the pure state $\vert \psis \rangle$, it is possible to saturate the equality $\vect{F}(\vect{\theta}){=}\vect{F_Q}(\vect{\theta})$ if and only if
\be \label{Matsumoto}
{\rm Im}[ \langle \partial_l \psis \vert \partial_m \psis \rangle] =0, \qquad \forall\, l,m.
\ee
To be more precise, the condition (\ref{Matsumoto}) is necessary and, if Eq.~(\ref{Matsumoto}) 
holds and the QFIM is invertible, Matsumoto proved that it is possible to constructs a set of projectors
for which $\vect{F}(\vect{\theta}){=}\vect{F_Q}(\vect{\theta})$ holds~\cite{MatsumotoJPA2002}.
Note that, for unitary transformations, Eq.~(\ref{Matsumoto}) becomes
$\langle \psis \vert [\hat{H}_l, \hat{H}_m] \vert \psis \rangle = 0$. 
Therefore, the (strong) commutativity condition $[\hat{H}_l, \hat{H}_m]=0$ for all $l,m$, implies Eq.~(\ref{Matsumoto}), 
while the saturation $\vect{F}(\vect{\theta}){=}\vect{F_Q}(\vect{\theta})$ is also possible for non-commuting generators~\cite{Baumgratz2016,MatsumotoJPA2002}.

Generally speaking, an experimental apparatus is set by a specific probe state, transformation and (projective) measurements
and it would be highly desirable to know whether -- within the specific setup -- it is possible to saturate the QFIM.
To this end, the weak commutativity theorem
is of little practical use: it only provides specific measurements for which $\vect{F}{=}\vect{F_Q}$ holds
that, however, might not be those implemented experimentally.
The saturation of the QFIM in an actual experiment is thus an open question in the literature, 
even for pure states and under the condition $[\hat{H}_l, \hat{H}_m]=0$ for all $l,m$.

In the following we provide three theorems giving necessary and sufficient conditions on projective measurements to
saturate $\vect{F}{=}\vect{F_Q}$ for arbitrary generator of phase encoding
(without assuming that $\vect{F_Q}^{-1}$ exists).
It is worth pointing out that if the FIM is invertible, then our conditions become necessary and sufficient 
to saturate the quantum Cramer-Rao bound with projective measurements.
We discuss the main consequences of our findings and the relation with existing results.
The proof of the theorems is detailed in appendix:

\textbf{Theorem 1 (Projective measurement orthogonal to the probe):}
Given a pure state $\vert \psis \rangle$ and
a set of projectors $\{ \vert \Upsilon_k \rangle \langle \Upsilon_k \vert \}_k$ on the state itself
($\vert \Upsilon_1 \rangle = \vert \psis \rangle$, for $k=1$) 
and on the orthogonal subspace ($\langle \Upsilon_k \vert \psis \rangle=0$, for $k\neq 1$),
the equality $\vect{F}(\vect{\theta}){=}\vect{F_Q}(\vect{\theta})$ holds if and only if
\be \label{result1}
\lim_{ \vect{\varphi} \to \vect{\theta} }
\frac{ {\rm Im}[ \langle \partial_l \psi_{\vect{\varphi}} \vert   \Upsilon_k \rangle \langle \Upsilon_k \vert \psi_{\vect{\varphi}} \rangle ] }
{ \vert \langle \Upsilon_k \vert  \psi_{\vect{\varphi}}  \rangle \vert } = 0,
\ee
for all $l=1,2,...,d$ and all $k\neq1$.

It is possible to demonstrate that Eq.~(\ref{result1}) is consistent with the weak commutativity condition (\ref{Matsumoto}), see appendix.
In the single parameter case, where $U(\theta)=e^{-i \hat{H} \theta}$ and $\hat{H}$ is a Hermitian operator, 
Eq.~(\ref{result1}) is always satisfied: it is thus always possible
to saturate the QFI by taking a measurement set made by the projector on the probe state $\vert \psis \rangle$ and
any set of projectors on the orthogonal subspace.
In general, this measurement is different from the set of projectors on the eigenstates of the symmetric logarithmic derivative (SLD), which also saturates the quantum Fisher information \cite{BraunsteinPRL1994}.
In Ref.~\cite{HumphreyPRL2013} it has been claimed that any projective measurement orthogonal to the probe state saturates the 
QFIM in the multi parameter case ($d>1$). 
We emphasize that only the orthogonal projective measurements which satisfy Eq.~(\ref{result1}) saturate the QFIM.
The condition~(\ref{result1}) is highly nontrivial:
it is easy to find examples of projective measurements that do not fulfill Eq.~(\ref{result1}), 
and for which $\vect{F}(\vect{\theta}) \neq \vect{F_Q}(\vect{\theta})$: an experimentally relevant example is discussed below.

If we assume that $\langle \Upsilon_k \vert  \partial_j \psis \rangle = 0$ does not hold for all $j=1,...,d$, then
[calculating the limit~(\ref{result1}), under the conditions of Theorem \#1]
$\vect{F}(\vect{\theta}) = \vect{F_Q}(\vect{\theta})$ if and only if 
\be \label{result1b}
{\rm Im} \big[ \langle \partial_l \psis  \vert \Upsilon_k \rangle \langle \Upsilon_k \vert \partial_m \psis \rangle \big]=0,
\ee
for all indices $l,m=1,2,...,d$ and all $k$.  
If $\langle \Upsilon_k \vert  \partial_j \psis \rangle = 0$ for all $j=1,...,d$ then it is possible to find necessary and sufficient conditions 
similar to Eq.~(\ref{result1b}) and involving higher order derivatives. 
A consequence of the above theorem is the following:

{\it Corollary \#1.} Given a probe pure state and unitary transformations, it is always possible to saturate $\vect{F}(\vect{\theta}){=}\vect{F_Q}(\vect{\theta})$
by a set of projectors given by the probe state itself and a suitable choice of vectors on the orthogonal subspace.

Here we explicitly construct optimal projectors saturating the QFIM.
We recall that the FIM at $\vts$ depends only on $\vert \psis \rangle$ and the $d$ vectors
$\{ \vert \partial_m \psis \rangle \}_{m=1,...,d}$.
In general, the states $\vert \partial_m \psis \rangle$ are not orthogonal to the probe.
To construct a set of projectors orthogonal to $\vert \psis \rangle$, 
we introduce the set of states
\be \label{lj}
\vert \omega_m \rangle \equiv \vert \partial_m \psis \rangle + \vert \psis \rangle \langle  \partial_m \psis  \vert \psis \rangle, 
\ee
for $m=1,...,d$. 
These satisfy
$\langle \psis \vert \omega_m \rangle = 2{\rm Re}[\langle  \partial_m \psis  \vert \psis \rangle] = \partial_m \langle \psis \vert \psis \rangle=0$ \cite{note2}.
The states~(\ref{lj}) are not orthogonal to each others, in general, and we can introduce the $d\times d$ Gram matrix
$\vect{\Omega}_{l,m}= \langle \omega_{l} \vert \omega_m \rangle = \langle \partial_l \psis \vert \partial_m \psis \rangle
+ \langle \partial_l \psis \vert \psis \rangle \langle \partial_m \psis \vert \psis \rangle$.
It should be noticed that, according to Eq.~(\ref{Matsumoto}), the saturation of the QFIM requires
${\rm Im} [ \vect{\Omega} ] = {\rm Im} [ \langle \partial_l \psis \vert \partial_m \psis \rangle ] =0$. \
We thus necessarily restrict to matrices $\vect{\Omega}$ that are real and symmetric. 
In particular, $\vect{F_Q} = 4 \vect{\Omega}$, see Eq.~(\ref{QuantumFisher}), 
and the QFIM is positive definite (and thus invertible) if and only if the vectors (\ref{lj}) are linearly independent.
We can construct, via the Gram-Schmidt process, for instance,
an orthogonal basis of the subspace $\mathcal{H} \setminus \vert \psis \rangle \langle \psis \vert$ as linear combinations of
states (\ref{lj})
\be
 \vert \Upsilon_k \rangle = \sum_{m=1}^d b_{m,k} \vert \omega_m \rangle,
\ee
with real coefficients $b_{m,k}$. 
With this choice, Eq.~(\ref{result1b}) is satisfied by the constructed set $\{\vert \Upsilon_k \rangle\}$ since
\be
\langle \partial_l \psis \vert \Upsilon_k \rangle = \sum_{m=1}^d \vect{\Omega}_{l,m} b_{m,k}
\ee
is real.
This concludes the proof of the corollary. 

The projectors of Theorem $\#$1 constitute a limited set: from an experimental perspective, a set of projectors orthogonal to a given probe state may not be easily available. In this respect, it is this interesting to derive statements for more general projectors.
We have:

\textbf{Theorem 2 (Projective measurement not orthogonal to the probe):}
Let us consider a probe pure state $\vert \psis \rangle$ and
a set of projectors $\{ \vert \Upsilon_k \rangle \langle \Upsilon_k \vert \}_k$ not orthogonal to the probe
(i.e. $\langle \Upsilon_k \vert \psis \rangle \neq 0$ for all $k$).
The equality $\vect{F}(\vect{\theta}){=}\vect{F_Q}(\vect{\theta})$ holds if and only if
\be \label{result2}
{\rm Im}[ \langle \partial_l \psis \vert \Upsilon_k \rangle \langle \Upsilon_k \vert \psis \rangle]
= \vert \langle \psis \vert \Upsilon_k \rangle \vert^2
{\rm Im}[ \langle \partial_l \psis \vert  \psis \rangle ],
\ee
for all $l=1,2,...,d$ and all $k$.

In the single parameter case, Eq.~(\ref{result2})
becomes ${\rm Re}[ \langle \psis \vert \hat{H} \vert \Upsilon_k \rangle \langle \Upsilon_k \vert \psis \rangle]
= \vert \langle \psis \vert \Upsilon_k \rangle \vert^2
\langle \psis \vert  \hat{H}\vert \psis \rangle$, for all $\Upsilon_k$.
This is precisely the necessary and sufficient conditions given in Ref.~\cite{WasakQIP2016} for the saturation of the quantum Fisher information
for the single parameter.
It is also possible to demonstrate that Eq.~(\ref{result2}) implies the weak commutativity condition (\ref{Matsumoto}),
see appendix for a demonstration.
Similarly as above we have:

{\it Corollary \#2.} Given a pure probe state and unitary transformations, there 
always exists a set of projectors non orthogonal to the probe which saturates $\vect{F}(\vect{\theta}){=}\vect{F_Q}(\vect{\theta})$.

We prove the corollary by constructing a set of projectors that satisfy Eq.~(\ref{result2}).
Restricting to the subspace $\mathcal{H}$, we can decompose the states $\vert \Upsilon_k \rangle$ in the basis given by $\vert \psis \rangle$ and
$\{ \vert \partial_m \psis \rangle \}_{m=1,...,d}$. We take
\be \label{condition_particular}
\vert \Upsilon_k \rangle = \sum_{m=1}^d b_{m,k} \vert \omega_m \rangle + b_{d+1,k} \vert \psis \rangle,
\ee
where $\vert \omega_n \rangle$ is given in Eq.~(\ref{lj}) and we require $b_{d+1,k} =  \langle \psis \vert \Upsilon_k \rangle \neq 0$ for all $k$.
Equation~(\ref{result2}) is fulfilled by taking real $b_{m,k}$ (for $m=1,2,...,d+1$)
and noticing that $ \langle \partial_l \psis \vert \omega_n \rangle$ is necessarily real to fulfill
Eq.~(\ref{Matsumoto}) \cite{note3}.
The real coefficients in Eq.~(\ref{condition_particular}) must be chosen such that $\sum_k \vert \Upsilon_k \rangle \langle \Upsilon_k |= \Eins$.
An orthonormal set that fulfills all these conditions can be constructed via the Gram-Schmidt process.

Finally, it is possible to combine the results of the two theorems and corollaries discussed above
and extend the saturation condition to an arbitrary set of projectors with elements that may be either orthogonal or not to the probe:

\textbf{Theorem 3 (General projective measurement):}
Consider a probe pure state $\vert \psis \rangle$ and
a set of projectors $\{ \vert \Upsilon_k \rangle \langle \Upsilon_k \vert \}_k$.
The equality $\vect{F}(\vect{\theta}){=}\vect{F_Q}(\vect{\theta})$ holds if and only if
Eq.~(\ref{result1})
is fulfilled for all indices $l,m$ and all $k$ for which $\langle \Upsilon_k \vert \psis \rangle = 0$,
and Eq.~(\ref{result2})
is fulfilled for all indices $l$ and all $k$ for which $\langle \Upsilon_k \vert \psis \rangle \neq 0$.

{\it Corollary \#3.} Given a pure probe state and unitary transformations, it is always possible to find a set of projectors satisfying the
theorem \#3 and thus saturate $\vect{F}(\vect{\theta}){=}\vect{F_Q}(\vect{\theta})$.

The proof of this statement immediately follows from Corollary \#2 by taking real coefficients
$b_{m,k}$ (for $m=1,...,d$), and $b_{d+1,k}$ either real and finite, or equal to zero. 
We point out that the projective measurement previously constructed by Matsumoto \cite{MatsumotoJPA2002} to 
saturated the equality $\vect{F}(\vect{\theta}){=}\vect{F_Q}(\vect{\theta})$ explicitly requires $\vect{F_Q}(\vect{\theta})$ to be invertible, 
a condition which is not general and not required in our case.
It is possible to show that the specific class of projective measurement used in the proof of the weak commutativity theorem 
satisfies the conditions of Theorem \#3.

\begin{figure}[b!]
\centering
\includegraphics[width=\columnwidth]{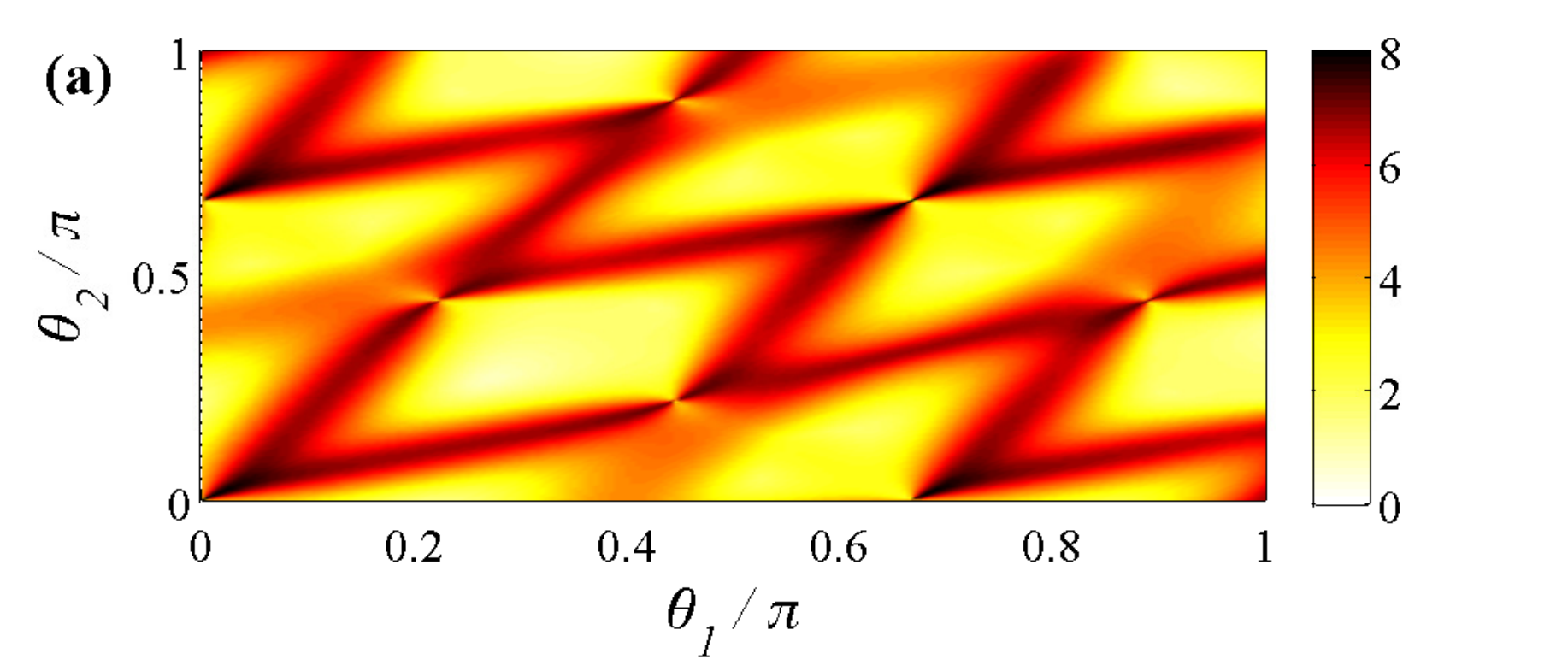}
\includegraphics[width=\columnwidth]{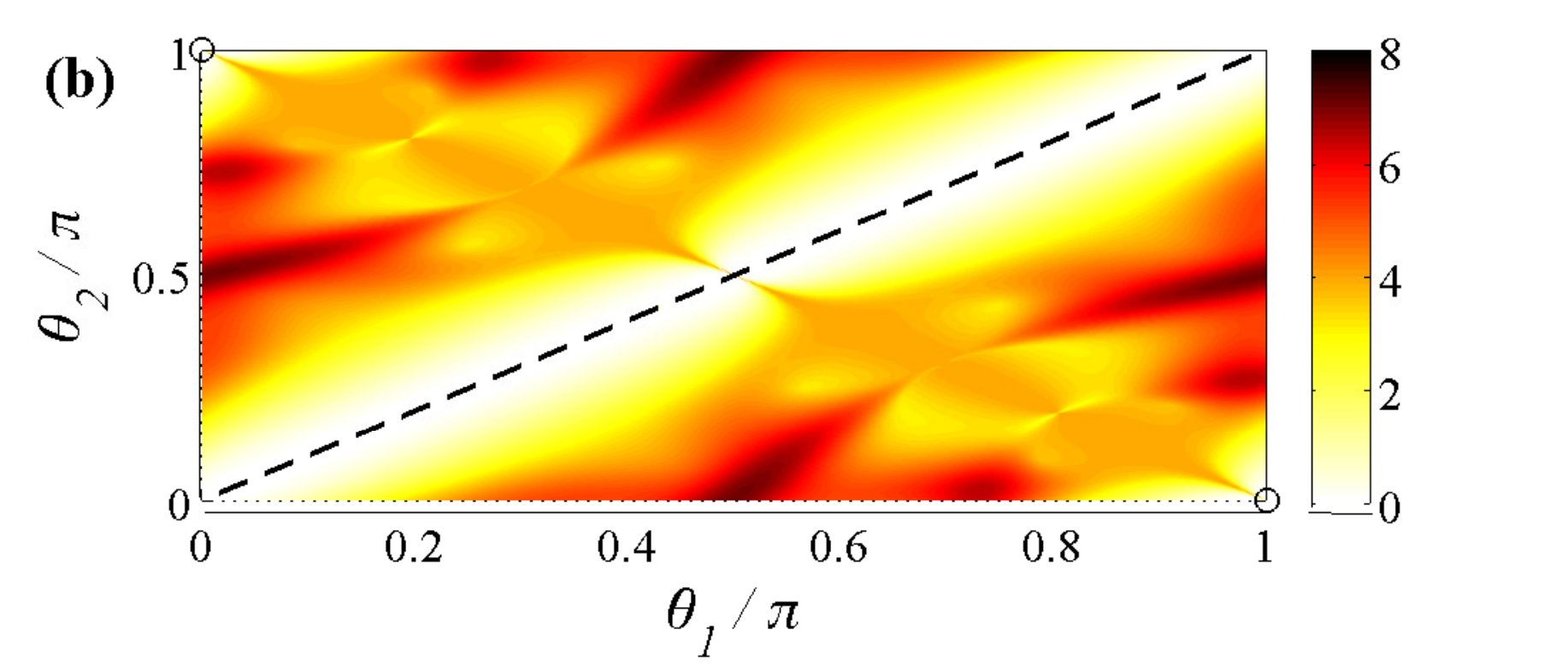}
\caption{$||\vect{F}-\vect{F_Q}||_2$ of a three-mode (a) and a four-mode (b) interferometer (see text and appendix)
for the estimation of $\vect{\theta}=(\theta_1,\theta_2)$. 
For the three-mode interferometer, there is no value of $\vect{\theta}$ where the theorems are satisfied.
Consistently, we find $||\vect{F}-\vect{F_Q}||_2>3/4$. 
For the four-mode interferometer, we can find values of $\vect{\theta}$ where the saturation conditions 
$\vect{F}=\vect{F_Q}$ discussed in the manuscript are fulfilled:
these are given by the dashed line and circles. }
\label{fig:MMZI}
\end{figure}

{\it Examples.}
Generally, optical and atomic interferometers use the measurement of the number of particles at the output ports in order to estimate the parameter(s).
Number counting realizes a projective measurement.
Using our results it is thus possible to tell whether or not, given a probe pure state and a phase encoding transformation, 
the FIM obtained with number counting saturates the QFIM.
Let us consider a $n$-mode Mach-Zehnder interferometer (MZI), which is a generalized version -- with $n$ arms -- of the 
common two-mode MZI \cite{GiovannettiNATPHOT2011, PezzeRMP2017}. 
This apparatus allows the simultaneous estimation of multiple (up to $n-1$) phases. 
Three- and four-mode (MZIs) are currently within reach of present technology \cite{SpagnoloSR2012,SpagnoloNC2013,Chaboyer2014}. 
We first discuss the three-mode case with a Fock state $| \psi \rangle = |1,1,1\rangle$ as input. 
The initial step of the interferometer is a three-mode splitter, a tritter, 
described by the unitary matrix $U^{(3)}_{j,k}=3^{-1/2}e^{{\rm i} 2 \pi/3 (1-\delta_{j,k})}$. 
Two optical phases $\theta_1$ and $\theta_2$ are inserted in two of the modes 
of the interferometer and are the parameters to be estimated simultaneously.
After phase encoding, the mode recombine at a second three-mode splitter, described by a unitary matrix $[U^{(3)}]^{-1}$.
Finally, photons in each mode are counted.
This measurement corresponds to a projection over all the possible states $|\Upsilon_k \rangle =|i,j,h\rangle$ with $i+j+h=3$.
We test the conditions (\ref{result1}) and (\ref{result2}) for different values of $\vect{\theta}=(\theta_1,\theta_2)$.
In particular, for $\vect{\theta}=(0,0)$, we have a projection over the probe state and over the orthogonal subspace.
We find that the condition (\ref{result1}) and (\ref{result2}) are not fulfilled. 
In Fig.~\ref{fig:MMZI}(a) we plot $||\vect{F}-\vect{F_Q}||_2$ by varying $\vect{\theta}$
(the norm $||\vect{F}-\vect{F_Q}||_2$ ranges from 0 to $||\vect{F_Q}||_2$ that is equal to 8 in this case.
We find that $||\vect{F}-\vect{F_Q}||_2>0$ ($||\vect{F}-\vect{F_Q}||_2>3/4$, in particular), consistent with the prediction of our theorems.
We have repeated the analysis for a four-mode interferometer, consisting of two cascaded quarters, in which 
a four photon $|1,1,1,1\rangle$  Fock state is injected. 
The quarters are optical devices represented by the unitary matrix $U^{(4)}_{j,k}=2^{-1} (-1)^{1-\delta_{j,k}}$. 
Again we consider the estimation of two phase and choose photon counting as measurement method. 
In this case, for certain values of $\vect{\theta}$ 
the conditions (\ref{result1}) and (\ref{result2}) are fulfilled and
the QFIM is saturated for the estimation of the two phases simultaneously: these are the values
$\theta_1 = \theta_2$ [dashed line in Fig.~\ref{fig:MMZI}(b)], and the points 
$\vect{\theta}=(0, \pi)$ and $\vect{\theta}=(\pi, 0)$ (circles).
Consistently, we find that $||\vect{F}-\vect{F_Q}||_2=0$ for these values of $\vect{\theta}$.

{\it Conclusions.}
While for single parameter estimation using pure probe states and unitary phase-encoding transformations
it is always possible to saturate the quantum Fisher information 
by projecting over the probe state and on the orthogonal subspace, this is not always the case for the estimation of multiple parameters.
Nontrivial conditions -- necessary and sufficient, as given in this paper -- must be satisfied in order to saturate the quantum Fisher information matrix.
We have also shown how to construct such optimal projectors and have tested the theory for an experimentally relevant configurations.
Finally, we recall that our conditions become necessary and sufficient for
the Cramer-Rao bound to saturate the quantum Cramer-Rao bound when the quantum Fisher information matrix is invertible.
Our results are a step forward to the theoretical understanding of multiparameter estimation:
the ability to saturate the QFIM is a key for the experimental design of future quantum imaging and multiparameter metrology devices. 

{\it Acknowledgements}.
M.B. is supported by a Rita Levi-Montalcini fellowship of MIUR. 
This work was supported by the ERC-Starting Grant 3D-QUEST (3D-Quantum Integrated Optical Simulation; grant agreement no. 307783): http://www.3dquest.eu, and by the H2020-FETPROACT-2014 Grant QUCHIP (Quantum Simulation on a Photonic Chip; grant agreement no. 641039): http://www.quchip.eu. AD is supported by the UK EPSRC (EP/K04057X/2) and the UK National Quantum Technologies Programme (EP/M01326X/1, EP/M013243/1).

\appendix

\newpage

\begin{widetext}

\section{Proof of theorem \#1.} 

Let us consider a projective set $\{ \vert \Upsilon_k \rangle \langle \Upsilon_k \vert \}_k$ 
where the element $k=1$ is the projection over the probe state 
$\vert \psis \rangle \equiv \hat{U}_{\vts} \vert \psi \rangle$, i.e. $\vert \Upsilon_1 \rangle = \vert \psis \rangle$,
and the other elements are chosen such that they are orthogonal to $\vert \psis \rangle$, 
i.e. $\langle \Upsilon_k \vert \psis \rangle=0$ for all $k\neq 1$. 
From Eq.~(3), the FIM at $\vect{\theta}$ is given by  
\be \label{Fisher1}
[\vect{F}(\vect{\theta})]_{l,m} = \lim_{\vtp \to \vect{\theta}} [\vect{F}(\vtp)]_{l,m}, 
\ee 
where 
\be \label{Fisher2} 
[\vect{F}(\vtp)]_{l,m}  = 
4 \, \frac{
{\rm Re}[ \langle \partial_l \psi_{\vtp} \vert \psis \rangle \langle \psis \vert \psi_{\vtp} \rangle ]  \, 
{\rm Re}[ \langle \partial_m \psi_{\vtp} \vert \psis \rangle \langle \psis \vert \psi_{\vtp} \rangle ] 
}{\vert \langle \psis \vert \psi_{\vtp} \rangle \vert^2}
+
4 \sum_{k\neq 1} 
\frac{
{\rm Re}[ \langle \partial_l \psi_{\vtp} \vert  \Upsilon_k \rangle \langle \Upsilon_k  \vert \psi_{\vtp}  \rangle]  \, 
{\rm Re}[ \langle \partial_m \psi_{\vtp} \vert  \Upsilon_k \rangle \langle \Upsilon_k \vert \psi_{\vtp} \rangle] 
}{ \vert \langle \Upsilon_k  \vert  \psi_{\vtp} \rangle \vert^2 }.
\ee
In the limit $\vtp \to \vect{\theta}$, we have $|\psi_{\vtp}\rangle \rightarrow \vert \psi_{\vts} \rangle \equiv\vert \psi_{s} \rangle$. 
The first term on the right side of Eq.~(\ref{Fisher2}) 
becomes $4 {\rm Re}[ \langle \partial_l \psis \vert \psis \rangle] {\rm Re}[ \langle \partial_m \psis \vert \psis \rangle]$
and it vanishes because $2{\rm Re}[ \langle \partial_j \psis  \vert \psis \rangle] 
= \partial_j \langle \psis  \vert \psis \rangle =0$ $\forall \, j$.
We thus have 
\be \label{Fisher3}
[\vect{F}(\vect{\theta})]_{l,m}  = 4 \lim_{\vtp \to \vect{\theta}}
\sum_{k\neq 1} 
\frac{
{\rm Re}[ \langle \partial_l \psi_{\vtp} \vert  \Upsilon_k \rangle \langle \Upsilon_k  \vert \psi_{\vtp}  \rangle]  \, 
{\rm Re}[ \langle \partial_m \psi_{\vtp} \vert  \Upsilon_k \rangle \langle \Upsilon_k \vert \psi_{\vtp} \rangle] 
}{ \vert \langle \Upsilon_k  \vert  \psi_{\vtp} \rangle \vert^2 }.
\ee
This limit is undetermined (0/0).
In the following we demonstrate that the inequality $\vect{F}(\vect{\theta}) \leq \vect{F_Q}(\vect{\theta})$ holds and find 
the necessary and sufficient condition for the saturation of the equality sign. 
From Eq.~(\ref{Fisher3}) we have 
\be \label{FishSum} 
\vect{u}^{\rm T} \vect{F}(\vect{\theta}) \vect{u} 
= \sum_{l,m=1}^d u_l  [\vect{F}(\vect{\theta})]_{l,m} u_m =  
4  \lim_{\vtp \to \vect{\theta}} \,  \sum_{k\neq1} \, 
\frac{\big( {\rm Re}[ \langle \Psi_{\vect{u}} \vert   \Upsilon_k \rangle \langle \Upsilon_k \vert \psi_{\vtp} \rangle ] \big)^2}
{ \vert \langle \Upsilon_k \vert \psi_{\vtp} \rangle \vert^2 }, 
\ee
where $\vert \Psi_{\vect{u}} \rangle \equiv \sum_{l=1}^d u_l \, \vert \partial_l \psi_{\vtp} \rangle$.
We now use $ {\rm Re}[x]^2=\vert x \vert^2 - {\rm Im}[x]^2$. 
In particular, 
$| \langle \Psi_{\vect{u}} \vert   \Upsilon_k \rangle \langle \Upsilon_k \vert \psi_{\vtp} \rangle |^2 / \vert \langle \Upsilon_k \vert \psi_{\vtp} \rangle \vert^2 = 
\vert \langle \Psi_{\vect{u}} \vert   \Upsilon_k \rangle \vert^2$.
We thus find
\beq
\vect{u}^{\rm T} \vect{F}(\vect{\theta}) \vect{u} &=& 
4 \lim_{\vtp \to \vect{\theta}} \, \sum_{k\neq1} \vert \langle \Psi_{\vect{u}} \vert   \Upsilon_k \rangle \vert^2  -
4 \lim_{\vtp \to \vect{\theta}} \,  \sum_{k\neq1} \, 
\frac{\big( {\rm Im}[ \langle \Psi_{\vect{u}} \vert   \Upsilon_k \rangle \langle \Upsilon_k \vert \psi_{\vtp} \rangle ] \big)^2}
{ \vert \langle \Upsilon_k \vert \psi_{\vtp} \rangle \vert^2 },  \\
 &=& 
 \vect{u}^{\rm T} \vect{F_Q}(\vect{\theta}) \vect{u} - 
 4 \lim_{\vtp \to \vect{\theta}} \,  \sum_{k\neq1} \, 
\frac{\big( {\rm Im}[ \langle \Psi_{\vect{u}} \vert   \Upsilon_k \rangle \langle \Upsilon_k \vert \psi_{\vtp} \rangle ] \big)^2}
{ \vert \langle \Upsilon_k \vert \psi_{\vtp} \rangle \vert^2 }. \label{Ineq1}
\eeq
To derive the second line we have used
$\lim_{\vtp \to \vect{\theta}} \, \vert \langle \Psi_{\vect{u}} \vert   \Upsilon_k \rangle \vert^2 = 
 \sum_{l,m=1}^d u_l  \langle \partial_l \psis\vert \Upsilon_k \rangle \langle \Upsilon_k \vert \partial_m \psis \rangle u_m$
and 
$\sum_{k\neq1} \vert \Upsilon_k \rangle \langle \Upsilon_k \vert = \Eins - \vert \psis \rangle \langle \psis \vert$.
The second term in Eq.~(\ref{Ineq1}) is always positive and bounded by $\vect{u}^{\rm T} \vect{F_Q}(\vect{\theta}) \vect{u}$.
Therefore, $\vect{F}(\vect{\theta}) \leq \vect{F_Q}(\vect{\theta})$, as expected, with equality if and only if the second term in Eq.~(\ref{Ineq1}) vanishes.
Since this is given by a sum (over $k\neq 1$) of positive terms, the equality is obtained if and only if each term of the sum vanishes 
for an arbitrary choice of $\vect{u}$. 
We thus conclude that $\vect{F}(\vect{\theta}) = \vect{F_Q}(\vect{\theta})$ holds if and only if 
\be \label{condition1} 
\boxed{ \lim_{\vtp \to \vect{\theta}}
\frac{ {\rm Im}[ \langle \partial_l \psi_{\vtp} \vert   \Upsilon_k \rangle \langle \Upsilon_k \vert \psi_{\vtp} \rangle ] }
{ \vert \langle \Upsilon_k \vert  \psi_{\vtp}  \rangle \vert } = 0, \qquad \forall \, l, \,\, \text{and} \, \, \forall \, k\neq 1.}
\ee
This is our most general condition and coincides with Eq.~(6) of the main paper.
The limit in Eq.~(\ref{condition1}) is undetermined (0/0) and to calculate it we 
proceed with a Taylor expansion of $\vert \psi_{\vtp}  \rangle$ and $\vert \partial_l \psi_{\vtp}  \rangle$.
To the leading order in $\delta \varphi = \varphi-\theta$, we have
\be \label{condition2}
\frac{ {\rm Im}[ \langle \partial_l \psi_{\vtp} \vert   \Upsilon_k \rangle \langle \Upsilon_k \vert \psi_{\vtp} \rangle ] }
{ \vert \langle \Upsilon_k \vert  \psi_{\vtp}  \rangle \vert } = 
\frac{ \sum_{j=1}^d {\rm Im}[ \langle \partial_l \psis \vert   \Upsilon_k \rangle \langle \Upsilon_k \vert \partial_j \psis \rangle ] \delta\varphi_j + O(\delta \varphi^2)}
{ \vert \sum_{j=1}^d  \langle \Upsilon_k \vert  \partial_j \psis  \rangle \delta \varphi_j + O(\delta \varphi^2) \vert }.
\ee
Excluding the case $\langle \Upsilon_k \vert  \partial_j \psis \rangle =0$ for all $j$, the limit exists and it is equal to zero if and only if 
\be  \label{NScond_0} 
\boxed{
{\rm Im} \big[ \langle \partial_l \psis  \vert \Upsilon_k \rangle \langle \Upsilon_k \vert \partial_m \psis \rangle \big]=0, 
\quad \quad \forall\,l,m, \,\, \forall k\neq 1. 
}
\ee
Clearly, the above condition is always satisfied when $l=m$.
The condition is thus nontrivial for $l\neq m$.  
If $\langle \Upsilon_k \vert  \partial_j \psis \rangle =0$ for all $j$, we need to consider the next order of the Taylor series 
in Eq.~(\ref{condition2}), giving conditions similar to Eq.~(\ref{NScond_0}) and involving higher order derivatives.

\subsection{Consistency with Matsumoto's condition}

Here we show that Eq.~(\ref{condition1}) implies Matsumoto's weak commutativity condition ${\rm Im} [ \langle \partial_l \psis  \vert \partial_m \psis \rangle ] =0$
for all $l,m=1,...,d$.
We have 
\be
 \lim_{\vtp \to \vect{\theta}} {\rm Im} \big[ \langle \partial_l \psi_{\vtp}  \vert \partial_m \psi_{\vtp} \rangle \big] = 
 \lim_{\vtp \to \vect{\theta}} \sum_{k\neq 1} {\rm Im} \big[ \langle \partial_l \psi_{\vtp}  \vert \Upsilon_k \rangle \langle \Upsilon_k \vert \partial_m \psi_{\vtp} \rangle \big],
\ee
where we have used $\sum_{k} \vert \Upsilon_k \rangle \langle \Upsilon_k \vert = \Eins$ and noticed that the term 
$\vert \Upsilon_1 \rangle = \vert \psis \rangle$ does not contribute to the sum since
${\rm Im} [ \langle \partial_l \psis  \vert \psis \rangle \langle \psis \vert \partial_m \psis \rangle]=0$.
We now multiply and divide by $\vert \langle \Upsilon_k \vert  \psi_{\vtp}  \rangle \vert^2$:
\beq
\lim_{\vtp \to \vect{\theta}} {\rm Im} \big[ \langle \partial_l \psi_{\vtp}  \vert \partial_m \psi_{\vtp} \rangle \big] 
&=& \lim_{\vtp \to \vect{\theta}} \sum_{k\neq 1}  \frac{ {\rm Im}[ \langle \partial_l \psi_{\vtp}  \vert \Upsilon_k \rangle \langle \Upsilon_k \vert \psi_{\vtp} \rangle
\langle \psi_{\vtp} \vert \Upsilon_k  \rangle \langle \Upsilon_k \vert \partial_m \psi_{\vtp} \rangle ] }
{ \vert \langle \Upsilon_k \vert  \psi_{\vtp}  \rangle \vert^2 } \\
&=&
\lim_{\vtp \to \vect{\theta}} \sum_{k\neq 1}  
\frac{ {\rm Im}[ \langle \partial_l \psi_{\vtp}  \vert \Upsilon_k \rangle \langle \Upsilon_k \vert \psi_{\vtp} \rangle]}
{\vert \langle \Upsilon_k \vert  \psi_{\vtp}  \rangle \vert}
\frac{{\rm Re}[\langle \psi_{\vtp} \vert \Upsilon_k  \rangle \langle \Upsilon_k \vert \partial_m \psi_{\vtp} \rangle ] }
{ \vert \langle \Upsilon_k \vert  \psi_{\vtp}  \rangle \vert } + \nonumber \\
&& \qquad \qquad \qquad + \frac{ {\rm Re}[ \langle \partial_l \psi_{\vtp}  \vert \Upsilon_k \rangle \langle \Upsilon_k \vert \psi_{\vtp} \rangle]}
{\vert \langle \Upsilon_k \vert  \psi_{\vtp}  \rangle \vert}
\frac{{\rm Im}[\langle \psi_{\vtp} \vert \Upsilon_k  \rangle \langle \Upsilon_k \vert \partial_m \psi_{\vtp} \rangle ] }
{ \vert \langle \Upsilon_k \vert  \psi_{\vtp}  \rangle \vert } \label{Matsu1}
\eeq
Because of Eq.~(\ref{condition1}), both the imaginary parts on the right-hand side of Eq.~(\ref{Matsu1}) vanish.
We thus conclude that 
${\rm Im} [ \langle \partial_l \psis  \vert \partial_m \psis \rangle ]  = 
\lim_{\vtp \to \vect{\theta}} {\rm Im} [ \langle \partial_l \psi_{\vtp}  \vert \partial_m \psi_{\vtp} \rangle ] = 0$.

\subsection{Single parameter case}

Here we show that Eq.~(\ref{condition1}) is always fulfilled in the single parameter case.
We expand Eq.~(\ref{condition1}) in Taylor series for $\varphi \approx \theta$
\be
\frac{ {\rm Im}[ \langle \partial \psi_{\varphi} \vert   \Upsilon_k \rangle \langle \Upsilon_k \vert \psi_{\varphi} \rangle ] }
{ \vert \langle \Upsilon_k \vert  \psi_{\varphi}  \rangle \vert } = 
\frac{\sum_{n=0}^{+\infty}\sum_{m=1}^{+\infty} \frac{1}{n!} \frac{1}{m!}
{\rm Im}[ \langle \partial^{(n+1)} \psis \vert   \Upsilon_k \rangle \langle \Upsilon_k \vert \partial^{(m)} \psis \rangle] (\delta \varphi)^{n+m}
}
{\vert \sum_{l=1}^{+ \infty} \frac{1}{l!} \langle \Upsilon_k \vert  \partial^{(l)} \psis  \rangle (\delta \varphi)^l \vert},
\ee
where $\delta \varphi = \varphi - \theta$ and we have taken into account that $\langle \Upsilon_k \vert \psis \rangle = 0$ for $k\neq 1$.
To collect all terms of the same order in $\delta \varphi$ in the numerator, we introduce 
$t = n+m$ and $r = (n-m)/2$:
\be \label{singpar}
\frac{ {\rm Im}[ \langle \partial \psi_{\varphi} \vert   \Upsilon_k \rangle \langle \Upsilon_k \vert \psi_{\varphi} \rangle ] }
{ \vert \langle \Upsilon_k \vert  \psi_{\varphi}  \rangle \vert } = 
\frac{\sum_{t=1}^{+\infty} \big( \sum_{r=-t/2}^{+t/2} \frac{1}{(t/2+r)!} \frac{1}{(t/2-r)!}
{\rm Im}[ \langle \partial^{(t/2+r+1)} \psis \vert   \Upsilon_k \rangle \langle \Upsilon_k \vert \partial^{(t/2-r)} \psis \rangle] \big) (\delta \varphi)^{t}
}
{\vert \sum_{l=1}^{+ \infty} \frac{1}{l!} \langle \Upsilon_k \vert  \partial^{(l)} \psis  \rangle (\delta \varphi)^l \vert}.
\ee
Let us suppose that the leading order in the denominator is $O(\delta \varphi)^o$
or, equivalently, $\langle \Upsilon_k \vert  \partial^{(l)} \psis  \rangle = 0$ for $l=1,2,...o-1$ and $\langle \Upsilon_k \vert  \partial^{(o)} \psis  \rangle \neq 0$.
Because of the null derivatives, the numerator is non-vanishing only if $t/2+r+1 \geq o$ and $t/2-r \geq o$, that implies $t \geq 2o -1$.
Therefor, for $o>1$, the the numerator in Eq.~(\ref{singpar}) 
is $O(\delta \varphi)^{o+1}$ or smaller, and the limit $\delta \varphi \to 0$ of the ratio converges to zero.
The case $o=1$ corresponds to $\langle \Upsilon_k \vert  \partial \psis  \rangle \neq 0$. In this case we have 
\be
\frac{ {\rm Im}[ \langle \partial \psi_{\varphi} \vert   \Upsilon_k \rangle \langle \Upsilon_k \vert \psi_{\varphi} \rangle ] }
{ \vert \langle \Upsilon_k \vert  \psi_{\varphi}  \rangle \vert } = 
\frac{{\rm Im}[ \langle \partial \psis \vert   \Upsilon_k \rangle \langle \Upsilon_k \vert \partial \psis \rangle] \delta \varphi + O(\delta \varphi)^2}
{\vert  \langle \Upsilon_k \vert  \partial \psis  \rangle \vert \delta \varphi + O(\delta \varphi)^2} = 0,
\ee
that vanishes because ${\rm Im}[ \langle \partial \psis \vert   \Upsilon_k \rangle \langle \Upsilon_k \vert \partial \psis \rangle]= 
{\rm Im} [\vert \langle \partial \psis \vert   \Upsilon_k \rangle \vert^2]=0 $.
We thus conclude that Eq.~(\ref{condition1}) is always satisfied in the single parameter case.

\section{Proof of theorem \#2.} 
Let us consider a set of projectors $\{ \vert \Upsilon_k \rangle \langle \Upsilon_k \vert \}_k$ not orthogonal to the probe state, 
i.e $\langle \Upsilon_k  \vert \psis \rangle \neq 0$ for all $k$.
The FIM, Eq.~(3), can be rewritten as 
\be \label{Fisher_notort1} 
[\vect{F}(\vect{\theta})]_{l,m}  = 
4 \sum_{k} 
\frac{
{\rm Re}[ \langle \omega_l \vert \Upsilon_k \rangle \langle \Upsilon_k \vert \psis \rangle]  \, 
{\rm Re}[ \langle \omega_m \vert \Upsilon_k \rangle \langle \Upsilon_k \vert \psis \rangle] 
}{ \vert \langle \Upsilon_k \vert \psis  \rangle \vert^2 },
 \ee
 where we have introduced 
 \be \label{dl_def} 
 \vert \omega_j \rangle \equiv \vert \partial_j \psis \rangle + \vert \psis \rangle \langle \partial_j \psis \vert \psis \rangle,
 \ee
and used
\be
{\rm Re}\big[ \langle \omega_j \vert \Upsilon_k \rangle \langle \Upsilon_k   \vert \psis \rangle \big] = 
{\rm Re}\big[ \langle \omega_j \vert \Upsilon_k \rangle \langle \Upsilon_k   \vert \psis \rangle \big] + 
\vert \langle \psis \vert \Upsilon_k  \rangle \vert^2 {\rm Re}\big[ \langle \psis \vert \partial_j \psis  \rangle \big] = 
{\rm Re}\big[ \langle \omega_j \vert \Upsilon_k \rangle \langle \Upsilon_k   \vert \psis \rangle \big],
\ee 
that holds since ${\rm Re}[ \langle \psis \vert \partial_j \psis  \rangle] =0$.
In the following we demonstrate that $\vect{F}(\vect{\theta}) \leq \vect{F_Q}(\vect{\theta})$ and find 
the necessary and sufficient condition (on the projective set acting on the Hilbert subspace $\mathcal{H}$) 
for the saturation of the equality sign. 
From Eq.~(\ref{Fisher_notort1}), we have 
\beq \label{Fisher_notort2}
\vect{u}^{\rm T} \vect{F}(\vect{\theta}) \vect{u} 
&=& \sum_{l,m=1}^d u_l [\vect{F}(\vect{\theta})]_{l,m} u_m =  
4 \sum_{k} 
\frac{\big( {\rm Re}[ \langle \Psi_{\vect{u}} \vert  \Upsilon_k \rangle \langle \Upsilon_k \vert \psis \rangle ] \big)^2}
{\vert \langle \Upsilon_k \vert \psis \rangle \vert^2}, \label{FishSum_new}
\eeq
where $\vert \Psi_{\vect{u}} \rangle \equiv \sum_{l=1}^d u_l \vert \omega_l \rangle$.
Similarly as above, we use ${\rm Re}[x]^2 = |x|^2 - {\rm Im}[x]^2$, giving 
\beq
\vect{u}^{\rm T} \vect{F}(\vect{\theta}) \vect{u}  &=& 4 \sum_k  \vert \langle \Psi_{\vect{u}} \vert  \Upsilon_k \rangle  \vert^2
- 4 \sum_k \frac{\big( {\rm Im}[ \langle \Psi_{\vect{u}} \vert  \Upsilon_k \rangle \langle \Upsilon_k \vert \psis \rangle ] \big)^2}
{\vert \langle \Upsilon_k \vert \psis \rangle \vert^2} \\
&=& \vect{u}^{\rm T} \vect{F_Q}(\vect{\theta}) \vect{u} - 4 \sum_k \frac{\big( {\rm Im}[ \langle \Psi_{\vect{u}} \vert  \Upsilon_k \rangle \langle \Upsilon_k \vert \psis \rangle ] \big)^2}
{\vert \langle \Upsilon_k \vert \psis \rangle \vert^2},
\eeq
where $\sum_k \vert \Upsilon_k \rangle \langle \Upsilon_k \vert = \Eins$  and $4  \sum_{k}
\langle  \Psi_{\vect{u}} \vert \Upsilon_k \rangle \langle \Upsilon_k \vert \Psi_{\vect{u}} \rangle
= 4 \langle  \Psi_{\vect{u}} \vert  \Psi_{\vect{u}} \rangle
= \vect{u}^{\rm T} \vect{F_Q}(\vect{\theta}) \vect{u}$.
Since $\vert \langle \Upsilon_k \vert \psis \rangle \vert^2 \neq 0$ for all $k$, 
the saturation of $\vect{F}(\vect{\theta}) = \vect{F_Q}(\vect{\theta})$ is obtained if and only if 
\be \label{condition_notort3}
{\rm Im}[ \langle \Psi_{\vect{u}} \vert  \Upsilon_k \rangle \langle \Upsilon_k \vert \psis \rangle ] =
\sum_l {\rm Im}\big[ \langle \omega_l  \vert \Upsilon_k \rangle \langle \Upsilon_k \vert \psis \rangle \big] u_l =0, \quad\quad \forall\,k.
\ee
Since this equality must be satisfied for all possible choices of $\vect{u}$, the necessary and sufficient condition is
\be
\boxed{ {\rm Im}[ \langle \omega_l \vert \Upsilon_k \rangle \langle \Upsilon_k \vert \psis \rangle ] = 0, \quad\quad \forall \, l,k}
\ee 
or, equivalently,
\be \label{condition_notort4} 
\boxed{ {\rm Im}\big[ \langle \partial_l \psis \vert \Upsilon_k \rangle \langle \Upsilon_k \vert \psis \rangle \big] 
- \big\vert \langle \psis \vert \Upsilon_k \rangle \big\vert^2 
 {\rm Im}\big[ \langle \partial_l \psis \vert  \psis \rangle \big]
= 0, 
\quad \quad \forall\,l, k. }
\ee
This concludes the demonstration of Theorem \#2.

\subsection{Consistency with Matsumoto's condition}

Here we show that Eq.~(\ref{condition_notort4}) implies Matsumoto's weak commutativity condition ${\rm Im} [ \langle \partial_l \psis  \vert \partial_m \psis \rangle ] =0$.
In other words, we want to show that, if 
\be \label{equivcond}
{\rm Im}[ \langle \omega_l \vert \Upsilon_k \rangle \langle \Upsilon_k \vert \psis \rangle ] = 
{\rm Im}[ \langle \omega_l \vert \Upsilon_k \rangle] {\rm Re} [\langle \Upsilon_k \vert \psis \rangle ] + 
{\rm Re}[ \langle \omega_l \vert \Upsilon_k \rangle] {\rm Im} [\langle \Upsilon_k \vert \psis \rangle ] = 0, \qquad \forall l,k,
\ee
then
${\rm Im}[ \langle \partial_l \psis  \vert \partial_m \psis \rangle ] 
= {\rm Im}[ \langle \omega_l  \vert \omega_m  \rangle ] =0$, $\forall l,m$.
We use the completeness condition $\sum_k \vert \Upsilon_k \rangle \langle \Upsilon_k \vert = \Eins$ and write  
\beq \label{Matsu}
{\rm Im}\big[ \langle \omega_l  \vert \omega_m  \rangle \big]
= \sum_k {\rm Im} [ \langle \omega_l  \vert \Upsilon_k \rangle \langle \Upsilon_k \vert \omega_m  \rangle ] 
=
\sum_k 
{\rm Im}\big[ \langle \omega_l  \vert \Upsilon_k \rangle \big] 
{\rm Re}\big[ \langle \Upsilon_k \vert \omega_m  \rangle \big] +
{\rm Re}\big[ \langle \omega_l  \vert \Upsilon_k \rangle \big] 
{\rm Im}\big[ \langle \Upsilon_k \vert \omega_m  \rangle \big]. \label{ReIm} 
\eeq
We are considering here projectors that are not orthogonal to the probe state, $\langle \Upsilon_k \vert \psis  \rangle \neq 0$. 
For a given projector (index $k$), there are three possible situations: 
({\it i}) ${\rm Re}[\langle \Upsilon_k \vert \psis \rangle] = 0$ and ${\rm Im}[\langle \Upsilon_k \vert \psis \rangle] \neq 0$, or
({\it ii}) ${\rm Re}[\langle \Upsilon_k \vert \psis \rangle] \neq 0$ and ${\rm Im}[\langle \Upsilon_k \vert \psis \rangle] = 0$, or 
({\it iii}) ${\rm Re}[\langle \Upsilon_k \vert \psis \rangle] \neq 0$ and ${\rm Im}[\langle \Upsilon_k \vert \psis \rangle] \neq 0$.
In the case ({\it i}), replacing ${\rm Re}[\langle \Upsilon_k \vert \psis \rangle] = 0$ into Eq.~(\ref{equivcond}), 
we find ${\rm Re}[ \langle \omega_j \vert \Upsilon_k \rangle]=0$ for all $j$.
This implies that the corresponding projector does not contribute to the sum in Eq.~(\ref{Matsu}).
The case ({\it ii}) is similar.
In the following we thus consider the case ({\it iii}), when both the real part and imaginary part of $\langle \Upsilon_k \vert \psis \rangle$ are finite. 
We multiply and divide into the sum of Eq.~(\ref{Matsu}) by ${\rm Re} [\langle \Upsilon_k \vert \psis \rangle ]$ and ${\rm Im} [\langle \Upsilon_k \vert \psis \rangle ]$:
\beq
{\rm Im}\big[ \langle \omega_l  \vert \omega_m  \rangle \big] &=& 
\sum_k 
{\rm Im}\big[ \langle \omega_l  \vert \Upsilon_k \rangle \big] {\rm Re} \big[\langle \Upsilon_k \vert \psis \rangle \big]
\frac{ {\rm Re} [ \langle \Upsilon_k \vert \omega_m  \rangle ] }{{\rm Re} [\langle \Upsilon_k \vert \psis \rangle ]}+
{\rm Re} [ \langle \omega_l  \vert \Upsilon_k \rangle ] {\rm Im} [\langle \Upsilon_k \vert \psis \rangle ]
\frac{{\rm Im} [ \langle \Upsilon_k \vert \omega_m  \rangle ]}{{\rm Im} [\langle \Upsilon_k \vert \psis \rangle ]}. 
\eeq
Using Eq.~(\ref{equivcond}) we obtain
\beq
{\rm Im}\big[ \langle \omega_l  \vert \omega_m  \rangle \big] &=&  
\sum_k {\rm Im}\big[ \langle \omega_l  \vert \Upsilon_k \rangle \big] {\rm Re} \big[\langle \Upsilon_k \vert \psis \rangle \big]
\bigg(  \frac{ {\rm Re} [ \langle \Upsilon_k \vert \omega_m  \rangle ] {\rm Im} [\langle \Upsilon_k \vert \psis \rangle ] 
-
{\rm Im} [ \langle \Upsilon_k \vert \omega_m  \rangle ] {\rm Re} [\langle \Upsilon_k \vert \psis \rangle] 
}
{{\rm Re} [\langle \Upsilon_k \vert \psis \rangle]  {\rm Im} [\langle \Upsilon_k \vert \psis \rangle] } \bigg) \nonumber \\
&=& \sum_k {\rm Im}\big[ \langle \omega_l  \vert \Upsilon_k \rangle \big] {\rm Re} \big[\langle \Upsilon_k \vert \psis \rangle \big]
\bigg(  \frac{ {\rm Re} [ \langle \omega_m \vert \Upsilon_k   \rangle ] {\rm Im} [\langle \Upsilon_k \vert \psis \rangle ] 
+
{\rm Im} [ \langle \omega_m \vert  \Upsilon_k   \rangle ] {\rm Re} [\langle \Upsilon_k \vert \psis \rangle] 
}
{{\rm Re} [\langle \Upsilon_k \vert \psis \rangle]  {\rm Im} [\langle \Upsilon_k \vert \psis \rangle] } \bigg) \nonumber \\
&=& 0,
\eeq
that vanishes because of Eq.~(\ref{equivcond}).
 
\section{Proof of theorem \#3.} 
We now consider a more general set of projectors $\{ \vert \Upsilon_k \rangle \langle \Upsilon_k \vert \}_k$. 
Some of the states $\vert \Upsilon_k \rangle$ are orthogonal to $\vert \psis \rangle$ and we label them 
with index $h$ ($\langle \Upsilon_{h} \vert \psis \rangle=0$). 
To calculate the contribution of these states to the FIM we follow the 
demonstration of Theorem \#1.
We label the other projectors -- those not orthogonal to the probe states -- with label $q$ ($\langle \Upsilon_{q} \vert \psis \rangle \neq 0$).
We have 
\beq \label{FisherTh3} 
[\vect{F}(\vect{\theta})]_{l,m} &=& 
 4 \lim_{\delta \vtp \to 0}
\sum_{h} 
\frac{
{\rm Re}[ \langle \partial_l \psi_{\vtp} \vert  \Upsilon_h \rangle \langle \Upsilon_h  \vert \psi_{\vtp}  \rangle]  \, 
{\rm Re}[ \langle \partial_m \psi_{\vtp} \vert  \Upsilon_h \rangle \langle \Upsilon_h \vert \psi_{\vtp} \rangle] 
}{ \vert \langle \Upsilon_h  \vert  \psi_{\vtp} \rangle \vert^2 } + \nonumber \\ 
&& +
4 \sum_{q} 
\frac{
{\rm Re}[ \langle \omega_l  \vert \Upsilon_q \rangle \langle \Upsilon_q \vert \psis \rangle]  \, 
{\rm Re}[ \langle \omega_m \vert \Upsilon_q \rangle \langle \Upsilon_q \vert \psis \rangle] 
}{ \vert \langle \Upsilon_q  \vert \psis \rangle \vert^2 },  
\eeq
where $\sum_h \vert \Upsilon_h \rangle \langle \Upsilon_h \vert + \sum_q \vert \Upsilon_q \rangle \langle \Upsilon_q \vert =\Eins$ and  
$\vert \omega_j  \rangle$ is given in Eq.~(\ref{dl_def}).
We have 
\beq 
\vect{u}^{\rm T} \vect{F}(\vect{\theta}) \vect{u}  &=& 4 \lim_{ \delta\vtp \to 0} \, \sum_{h} \vert \langle \Psi_{\vect{u}} \vert   \Upsilon_h \rangle \vert^2  
- 4 \lim_{ \delta\vtp \to 0} \,  \sum_{h} \, 
\frac{\big( {\rm Im}[ \langle \Psi_{\vect{u}} \vert   \Upsilon_h \rangle \langle \Upsilon_h \vert \psi_{\vtp} \rangle ] \big)^2}
{ \vert \langle \Upsilon_h \vert \psi_{\vtp} \rangle \vert^2 } \nonumber \\
&& 
+ 4 \sum_q  \vert \langle \Psi_{\vect{u}} \vert  \Upsilon_q \rangle  \vert^2
- 4 \sum_q \frac{\big( {\rm Im}[ \langle \Psi_{\vect{u}} \vert  \Upsilon_q \rangle \langle \Upsilon_q \vert \psis \rangle ] \big)^2}
{\vert \langle \Upsilon_q \vert \psis \rangle \vert^2} \\
&=& \vect{u}^{\rm T} \vect{F_Q}(\vect{\theta}) \vect{u} - 4 \lim_{ \delta\vtp \to 0} \,  \sum_{h} \, 
\frac{\big( {\rm Im}[ \langle \Psi_{\vect{u}} \vert   \Upsilon_h \rangle \langle \Upsilon_h \vert \psi_{\vtp} \rangle ] \big)^2}
{ \vert \langle \Upsilon_h \vert \psi_{\vtp} \rangle \vert^2 }
- 4 \sum_q \frac{\big( {\rm Im}[ \langle \Psi_{\vect{u}} \vert  \Upsilon_q \rangle \langle \Upsilon_q \vert \psis \rangle ] \big)^2}
{\vert \langle \Upsilon_q \vert \psis \rangle \vert^2},
\eeq
where
$ \sum_{h}
\langle  \Psi_{\vect{u}} \vert \Upsilon_h \rangle \langle \Upsilon_h \vert \Psi_{\vect{u}} \rangle
+ 4  \sum_{q}
\langle  \Psi_{\vect{u}} \vert \Upsilon_q \rangle \langle \Upsilon_q \vert \Psi_{\vect{u}} \rangle
= 4 \langle  \Psi_{\vect{u}} \vert  \Psi_{\vect{u}} \rangle
= \vect{u}^{\rm T} F_Q(\vect{\theta}) \vect{u}$.
The equality $\vect{F}(\vect{\theta})=\vect{F_Q}(\vect{\theta})$ is recovered if and only if 
\begin{subequations}
\begin{empheq}[box=\widefbox]{align}
 & \lim_{ \delta\vtp \to 0 }
\frac{ {\rm Im}[ \langle \partial_l \psi_{\vtp} \vert   \Upsilon_h \rangle \langle \Upsilon_h \vert \psi_{\vtp} \rangle ] }
{ \vert \langle \Upsilon_h \vert  \psi_{\vtp}  \rangle \vert } = 0, \qquad \forall \, l, \,\,\, \forall \, h, \\
 & {\rm Im}\big[ \langle \partial_l \psis \vert \Upsilon_q \rangle \langle \Upsilon_q \vert \psis \rangle \big] 
- \big\vert \langle \psis \vert \Upsilon_q \rangle \big\vert^2 
 {\rm Im}\big[ \langle \partial_l \psis \vert  \psis \rangle \big]
= 0, 
\quad \quad \forall\,l, \,\,\, \forall \, q.
\end{empheq}
\end{subequations}

\section{Example: the multimode Mach-Zehnder interferometers}

We discuss in details the example provided in the main text. 
Let us consider a $n$-mode interferometer, composed by two cascaded multiport splitters. 
The parameters to be estimated are a set of $d=n-1$ optical phases $\vect{\theta} = (\theta_{1}, \ldots, \theta_{n-1})$, 
whose action on the probe state is described by a unitary evolution $U(\vect{\theta})$.

In the three-mode case, the modes of the interferometer transform according to the unitary $U(\vect{\theta}) = [U^{(3)}]^{-1} U_{\rm ps}(\vect{\theta}) U^{(3)}$, where 
$U^{(3)}$ is a tritter and $U_{\rm ps}(\vect{\theta})$ provides a shift of phase of two modes with respect to the third mode:
\begin{equation}
U^{(3)} = \frac{1}{\sqrt{3}} \begin{pmatrix} 
1 & e^{{\rm i} 2 \pi/3} & e^{{\rm i} 2 \pi/3} \\ 
e^{{\rm i} 2 \pi/3} & 1 & e^{{\rm i} 2 \pi/3} \\
e^{{\rm i} 2 \pi/3} & e^{{\rm i} 2 \pi/3} & 1 
\end{pmatrix},
\quad {\rm and} \quad 
U_{\rm ps}(\vect{\theta}) = \begin{pmatrix} 
e^{{\rm i} \theta_1} & 0 & 0 \\ 
0 & e^{{\rm i} \theta_2} & 0 \\
0 & 0 & 1 
\end{pmatrix}.
\end{equation}
The system can be adopted to estimate two optical phases $\vect{\theta} = (\theta_{1}, \theta_{2})$. 
A direct analytical calculation of the QFIM gives
\be
\vect{F_Q} = \frac{8}{3} 
\begin{pmatrix} 
2 & -1 \\ 
-1 & 2 \\
\end{pmatrix}
\ee
We now test our conditions for the saturation of the equality $\vect{F}(\vect{\theta}) = \vect{F_Q}$, when 
the input is a $ \vert \psi \rangle = \vert 1,1,1 \rangle$ Fock state, and for photon-counting measurements.
This measurement strategy corresponds to a projection over Fock states $\vert \Upsilon_{k} \rangle = \vert i,j,h \rangle$, with $i+j+h=3$:
\begin{equation}
\begin{aligned}
\vert \Upsilon_{1} \rangle &= \vert 1,1,1\rangle; \quad \vert \Upsilon_{2} \rangle = \vert 2,1,0\rangle; \quad \vert \Upsilon_{3} \rangle = \vert 2,0,1\rangle; \quad \vert \Upsilon_{4} \rangle = \vert 1,2,0\rangle; \quad \vert \Upsilon_{5} \rangle = \vert 1,0,2\rangle; \\
\vert \Upsilon_{6} \rangle &= \vert 0,2,1\rangle; \quad \vert \Upsilon_{7} \rangle = \vert 0,1,2\rangle; \quad \vert \Upsilon_{8} \rangle = \vert 3,0,0\rangle; \quad \vert \Upsilon_{9} \rangle = \vert 0,3,0\rangle; \quad \vert \Upsilon_{10} \rangle = \vert 0,0,3\rangle.
\end{aligned}
\end{equation}
For $\vect{\theta} = (0,0)$, the state $\vert \Upsilon_{1} \rangle$ corresponds to the projector over
$\vert \psi_{s} \rangle = \vert \psi \rangle$.
We can then test condition (6) for this choice of a projective measurement.
A direct calculation of $C_k={\rm Im} [ \langle \partial_1 \psis  \vert \Upsilon_k \rangle \langle \Upsilon_k \vert \partial_2 \psis \rangle ]$
gives $C_k=1/(3\sqrt{3})$ for $k=2,5,6$, $C_k=-1/(3\sqrt{3})$ for $k=3,4,7$ and $C_k=0$ for $k=1,8,9,10$:
we thus conclude that the condition (6) for the saturation of the QFIM is not satisfied.
Indeed, using Eq.~(\ref{Fisher3}), the FIM at $\vect{\theta}=(0,0)$ is
\be
\vect{F}(0) = \frac{4}{3} \begin{pmatrix} 
1 & 1 \\ 
1 & 1 \\
\end{pmatrix}.
\ee

A similar analysis can be performed for the four-mode interferometer. 
This interferometer can be adopted for the estimation of three phases. 
Here, to compare with the three-mode interferometer, we consider the estimation of two phases. 
In this case, $U(\vect{\theta}) = [U^{(4)}]^{-1} U_{\rm ps}(\vect{\theta}) U^{(4)}$, where $U^{(4)}$ is a quarter:
\begin{equation}
U^{(4)} = \frac{1}{2} \begin{pmatrix}
1 & -1 & -1 & -1\\
-1 & 1 & -1 & -1\\
-1 & -1 & 1 & -1\\
-1 & -1 & -1 & 1
\end{pmatrix}
\quad {\rm and} \quad 
U_{\rm ps}(\vect{\theta}) = \begin{pmatrix} 
e^{{\rm i} \theta_1} & 0 & 0 & 0\\ 
0 & e^{{\rm i} \theta_2} & 0 & 0\\
0 & 0 & 1 &0 \\
0 & 0 & 0 & 1 
\end{pmatrix}.
\end{equation}
We take $\vert 1,1,1,1 \rangle$ as input. 
The QFIM is 
\be
\vect{F_Q} = 2 \begin{pmatrix} 
3 & -1 \\ 
-1 & 3 \\
\end{pmatrix}.
\ee
We consider a projective measurement over Fock states 
$\vert \Upsilon_{k} \rangle = \vert i,j,g,h \rangle$ with $i+j+g+h=4$.
For $\vect{\theta} = (0,0)$, the measurement consists of a projection over the probe state and over the orthogonal subspace.
At variance with the three-mode case, we evaluate $C_{k,l,m}={\rm Im} [ \langle \partial_l \psis  \vert \Upsilon_k \rangle \langle \Upsilon_k \vert \partial_m \psis \rangle ]$
and find $C_{k,l,m}=0$ for all $k\neq 1$, $l,m=1,\ldots, 3$: the condition (6) is satisfied and 
the equality $\vect{F}(0) = \vect{F_Q}$ is saturated.
The saturation has been further tested for all values of $\vect{\theta}$, se text.

\end{widetext}


\begin{thebibliography}{10}
\bibitem{note2}
Notice that $\vert \omega_m \rangle = \hat{L}_m \vert \psi_s \rangle/2$, 
where $\hat{L}_m = 2( \vert \partial_m \psis\rangle \langle \psis \vert + \vert \psis \rangle \langle \partial_m \psis \vert)$ is the symmetric logarithmic derivative defined as 
$2\partial_m ( \vert \psis \rangle \langle \psis \vert ) = \hat{L}_m \vert \psis \rangle \langle \psis \vert  + \vert \psis \rangle \langle \psis \vert  \hat{L}_m$.
We have $4{\rm Re}[ \langle \omega_l \vert \omega_m \rangle] = [\vect{F}_Q]_{l,m}$.
\end{thebibliography}
\end{document}